\documentstyle[twocolumn,aps]{revtex}

\input{epsf}

\def\beq{\begin{equation}}
\def\eeq{\end{equation}}

\begin{document}


\title      {  The power of surrogate data testing with respect to
                 non-stationarity  }
\author {       J. Timmer }

\address{ Physics Department, University of Freiburg, Hermann-Herder-Str. 3, 
	D-79104 Freiburg, Germany}

\maketitle

\begin {abstract}
Surrogate data testing is a method frequently applied to evaluate
the results of nonlinear time series analysis.
Since the null hypothesis tested against is 
a linear, gaussian, stationary stochastic process a positive outcome may
not only result from an underlying nonlinear or even chaotic system,
but also from e.g.~a non-stationary linear one.
We investigate the power of the test against non-stationarity.
\end {abstract}

\pacs{05.45.+b, 02.50.Fz}
\section*{I. INTRODUCTION }
The field of non-linear dynamics introduced the fascinating idea that
an apparently random
behavior of a time series might have been generated by a low dimensional
deterministic system \cite{may76}. Based on the notions of chaos theory,
different algorithms have been invented to infer if an observed
time series is a realization of a chaotic system, e.g.~the
estimation of the largest Lyapunov-exponent \cite{wolf85}, the
correlation dimension \cite{grassberger83a} and nonlinear prediction
\cite{sugihara90}. There is hope to gain deeper insights in complex
systems like those from biology and physiology by applying these methods.

However, the application of these methods to a finite, often noisy
set of measured data is not straightforward, see 
e.g.~\cite{theiler92,rapp93b,glass93,jedynak94,kantz95} and 
references therein. For example, in order to claim a finite, fractal 
correlation dimension, a scaling region of sufficient
length has to be established. Determining this scaling region by eye
or some algorithm may lead to an erroneous evidence of chaotic
behavior. In order to evaluate the analysis, it
has become popular to apply the method of surrogate data \cite {theiler92}. 
Therefore, data are generated which have the same linear statistical
properties as the original data but not the possible nonlinear ones.
For many realizations of these data, the same algorithm as to the
original data is applied. A significant difference between
the distribution of the nonlinear feature for the surrogate data and the
original data is taken as an indication that
the process underlying the original data is deterministic
\cite{schiff94}, nonlinear \cite{stam95,mrowka96,guzzetti96} or even chaotic
\cite{yamamoto93,yip95,pradhan96}.

The explicit null hypotheses of surrogate data testing for linearity is
that the data were generated by a linear, stochastic, gaussian 
stationary process, including a possible invertible nonlinear
observation function. Thus, a rejection of this hypothesis does 
not necessarily mean that the data come from a chaotic, i.e.~some kind of 
stationary, nonlinear deterministic, process. They might also originate
from a nonlinear stochastic or even simply from a linear,
stochastic, non-stationary process. In this paper, we 
investigate the power of surrogate data testing against
non-stationarity.
As nonlinear feature we use the correlation dimension. The behavior of 
correlation dimension estimates has been investigated for
the $1/f^{\alpha}$, $\alpha \ge 1$ type of linear non-stationarity 
\cite{osborne89,theiler91}. For physiological data, such $1/f$ 
behavior has been observed in heart rate \cite{goldberger90}.
Often, physiological data are characterized by some kind of
oscillatory behavior like EEG, hormone secretion, breathing or
tremor. For such data, types of non-stationarity introducing
some time dependency of the oscillating dynamics, e.g~a modulation of
frequency or amplitude, seems to be a natural violation of 
the null hypothesis. 

If the process is linear and the time dependency of the parameters,
and thus, the autocovariance function is periodically in 
time, these processes are called cyclostationary \cite {garnder90}. 
Many other types of non-stationarity in oscillatory processes are 
imaginable. We choose cyclostationary processes because they 
allow in simple way for a parametric violation of the null hypothesis.
Formally, these processes can be expressed as higher dimensional
autonomous non-linear stochastic processes. A special version of 
surrogate data testing acting on segments of the data has been 
suggested to analyze such data \cite{kaplan97}. 

In the next section, we informally discuss
the class of cyclostationary processes and introduce the two specific 
examples we use in Section III to investigate the power of surrogate 
data testing with respect to these types of non-stationarity.
\section*{II. Cyclostationary Processes}
The parameters $a_i$ and $\sigma^2$ of a linear stochastic
autoregressive (AR) process $x(t)$ : 
\beq \label{ar_process}
 x(t) = \sum_{i=1}^p a_1 x(t-p) + \epsilon (t), \quad \epsilon (t)
 \sim {\cal{N}}(0,\sigma^2)
\eeq
determine the autocovariance function  $R(\tau)$  :
\beq
  R(\tau) = < x(t) x(t+\tau)> \quad .
\eeq
The spectrum $ S(\omega) $ is given as Fourier transform of the 
autocovariance function~:
\beq
  S(\omega) = \sum e^{-i\omega \tau}  R(\tau) \quad .
\eeq
A possible first step to non-stationarity is to define a time
 dependent spectrum
$ S(t,\omega) $ and, correspondingly, a time dependent autocovariance
function $ R(t,\tau) $:
\beq
  R(t,\tau) = < x(t) x(t+\tau)> \quad .
\eeq
A cyclostationary process of periodicity $L$ is defined by:
\beq \label{cyclo_def}
  R(t,\tau) = R(t+L,\tau) \quad . 
\eeq
For the AR process of Eq. (\ref{ar_process}) this means that
the parameters $a_i$ and $\sigma^2$ may change periodically.

As process satisfying the null hypothesis of surrogate
data testing for linearity, we chose an autoregressive (AR)
process of order two:
\beq 
 x_{t} =  a_1 x_{t-1}  + a_2 x_{t-2}  + \epsilon_{t}, 
	\quad 	\epsilon_{t} \sim {\cal{N}}(0,\sigma^2) \, .
\eeq

In terms of physics, AR processes can be interpreted as a combination 
of linear relaxators and linear damped oscillators driven by noise.
For an AR process of order two which describes
a damped oscillator, the parameters are related to the relaxation
time $\tau$ and period $T$ by :
\begin{eqnarray}
  a_1 & = & 2 \cos \left(2\pi/T\right) \, 
		\exp \,(-1/\tau)\label{a1}\\ 
  a_2 & = & - \exp \,(-2/\tau) \label{a2}  \qquad \qquad  .
\end{eqnarray}
The variance of the process Var($x_t$) is given by :
\beq \label{var_ar2}
\mbox{Var}(x_t)= \frac{ \sigma^2  }  {  1-a_1^2 -a_2^2 - 
		\frac{2a_1^2a_2 }{1-a_2} } \quad .
\eeq

\begin{figure}
\epsfxsize=8.4cm
 \epsfbox{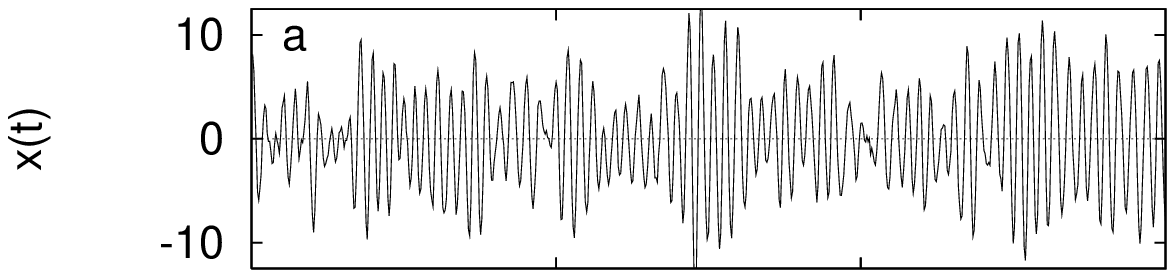}
\par\vspace{-0.5cm}
\epsfxsize=8.4cm
 \epsfbox{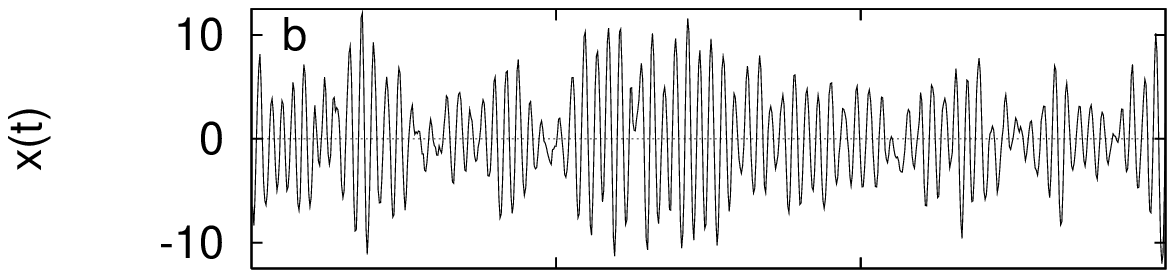}
\par\vspace{-0.5cm}
\epsfxsize=8.4cm
 \epsfbox{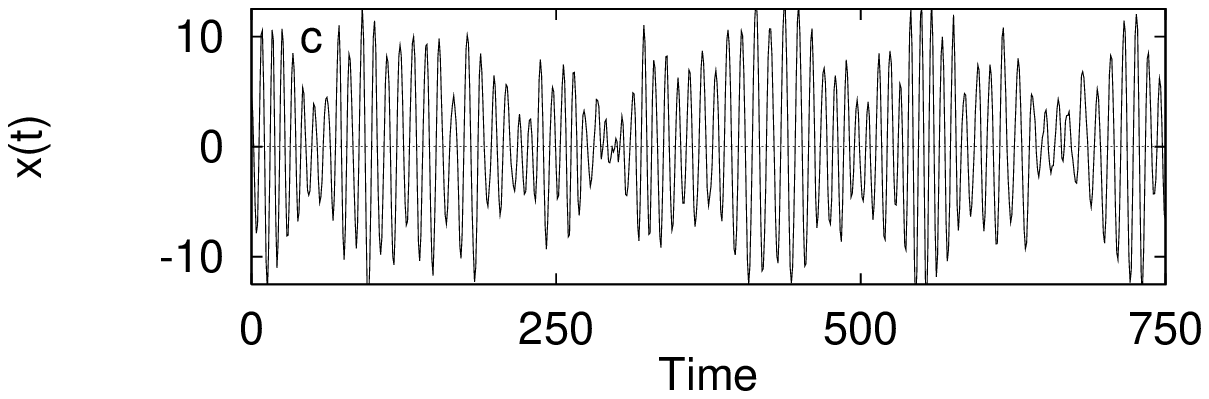}
\caption{\label{ar2_dat} Realizations of the processes investigated. 
	(a) AR2 process satisfying the null hypothesis. 
	(b) Amplitude modulated process with modulation depth of 0.3.
	(c) Period modulated process, relative amplitude of
	modulation is 15\%. }
\end{figure}

We choose an AR2 process with $T=10$, $\tau=50$ and $\sigma=1$ as
process  $ x_0(t) $ that satisfies the null
hypothesis. Fig.~\ref{ar2_dat}a displays a realization of this process.

\begin{figure}
\epsfxsize=8.4cm 
\epsfbox{  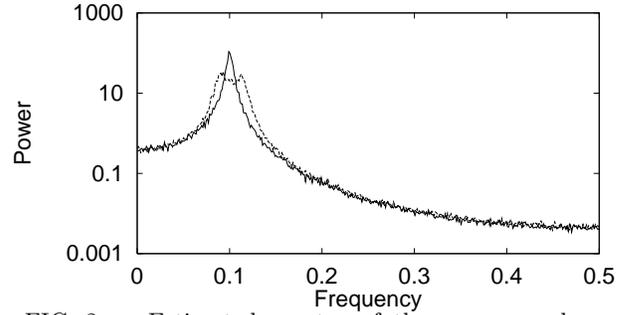 }
\caption{\label{ar2_spe} Estimated spectra of the processes shown in
  Fig.~\ref{ar2_dat}. The spectra corresponding to Fig.~\ref{ar2_dat}a
  and b are not distinguishable (solid line). Period modulated process
	(dashed line).}
\end{figure}

The oscillatory behavior with a mean 
period of 10 time steps is clearly visible as well as the natural 
variability of period and amplitude. Fig.~\ref{ar2_spe} 
(solid line) shows the estimated spectrum of the process. 
The 
spectrum was estimated by averaging 100 periodograms, i.e.~the 
squared absolute value of the Fourier transform of the data.

A broad peak, typical for a stochastically driven linear damped
oscillator can be seen.
Based on Eqs.~(\ref{a1},\ref{a2},\ref{var_ar2}) we now introduce two
parameterized violations of this stationary, linear, stochastic
process in order to investigate the power of surrogate data testing
with respect to non-stationarity.

For the first violation of stationarity in the frame of
cyclostationary processes, we choose a simple amplitude
modulation, corresponding by Eq.(\ref{var_ar2}) to a periodicity of
the variance of the driving noise. Based on the stationary AR2
process $ x_0(t) $, the amplitude modulated process $ x_{amp}(t) $ 
is given by :
\beq \label{process1}
x_{amp}(t) = (1+\text{Mod}_{\text{amp}} \sin (2\pi/T_{mod} \,t)) \,
x_0(t) \quad . 
\eeq
$\text{Mod}_{\text{amp}}$, the modulation depth, parameterizes the
violation of the null hypothesis. $ T_{mod} $ determines the
modulation period. Fig.~\ref{ar2_dat}b displays
a realization of this process with $ T_{mod}=250  $ and $
\text{Mod}_{\text{amp}}=0.3 $ for three periods of the modulation.
Compared to Fig.~\ref{ar2_dat}a, the non-stationarity is hardly visible.
Due to the long modulation period compared
to the period of the process, its spectrum is not distinguishable from
that of the stationary process in Fig.~\ref{ar2_spe}.

For the second violation of stationarity, we chose a modulation of
the period $T$ of the AR2 process
with period $ T_{mod} $ and amplitude $ \text{Mod}_T $ around the mean
period $ T_{mean}=10 $. This leads to a time dependency of the
parameter $ a_1 $ of the AR2 process :
\begin{eqnarray}
  T(t) & = &T_{mean} \,+ \, \text{Mod}_T \sin (2\pi/T_{mod} \, t)
                          \label{process2_1} \\
  a_1(t) & = & 2 \cos \left(2\pi/T(t)\right) \, 
        \exp \,(-1/\tau) \label{process2_2} \qquad .
\end{eqnarray}
$\text{Mod}_T$ parameterizes the violation of the null hypothesis.
According to Eq.~(\ref{var_ar2}), the time dependency of $ a_1(t) $
causes a time dependency of the variance of the process. The effect of
a changing variance is already covered by the first process, 
Eq.~(\ref{process1}). To investigate only the effect of a changing 
period of the process here, we use Eq.~(\ref{var_ar2}) to adjust the 
variance $ \sigma^2(t) $ of the
driving noise such that the variance of the process is constant : 

\begin{eqnarray} 
\sigma^2(t) & = &  \frac{ \sigma^2  }  {  1-a_1^2 -a_2^2 - 
                      \frac{2a_1^2a_2}{1-a_2}} \\
 & & \times \left( 1-a_1(t)^2 -a_2^2 - \frac{2a_1(t)^2a_2 }{1-a_2}
                      \right)  \quad ,
\end{eqnarray} 
where $a_1$ and $\sigma^2$ denote the parameters of the  process
$x_0(t)$ satisfying the null hypothesis. Fig.~\ref{ar2_dat}c displays
a realization of this process with $ T_{mod}=250 $ and 
$ \text{Mod}_T=1.5 $. Again, compared to Fig.~\ref{ar2_dat}a, the 
non-stationarity is hardly visible.
Fig.~\ref{ar2_spe} (dashed line) shows the estimated
spectrum of the process. The spectrum shows two peaks at the
corresponding frequencies due to the specific type of modulation chosen.

\section*{III. Power of the test} \label{sec_power}

As nonlinear feature to investigate the power of surrogate data testing
against the two violations of stationarity we use the correlation
dimension. The phase space is reconstructed by delay embedding.
The delay is chosen equal to the lag at which the autocorrelation 
function first crosses zero.

The correlation dimension $D_{2}$ is defined by:
 \begin {equation} \label {D2}
 D_{2}= \lim_{r \rightarrow 0} \frac { d\, \ln C(r) } {d\, \ln r } \quad ,
 \end {equation}
where C(r), the correlation integral, is given by:
 \begin {equation} \label {Cr}
 C(r) = \text{const} \sum _{i=1} ^{N-\mu} \sum_{j = i +\mu}^{N}
      \Theta (r-|\vec{x}(i)-\vec{x}(j)|) \quad ,
 \end {equation}
including the Theiler correction $ \mu $ \cite{theiler86} which
 we chose equal to the mean period, i.e.~10 time steps.
The canonical procedure to establish a finite correlation
dimension is to show the existence of a scaling region for small $r$ 
where Eq.~(\ref{D2}) holds and stays constant
for a high enough embedding dimensions. 
For all processes investigated here, the true correlation dimension
is infinity. Following the idea of surrogate data
testing, we fix an algorithm to obtain a finite value from
the correlation integral and look for differences to the original 
data. Therefore, we apply Theiler and Lookman's ''rule of five'' 
chord estimator \cite{theiler93} and chose their
$R_0$ equal to the standard deviation of the data. 
For such a large $R_0$ we do not examine the small scale behavior of
Eq.~(\ref{D2}) anymore. We are aware that we should not call this
quantity correlation dimension anymore. It has been termed 
''dimensional complexity'' \cite{pritchard92}. 

The surrogate data are produced by the FFT algorithm \cite{theiler92}.
For each degree of violation of the null hypothesis 50 independent 
surrogate data sets of length 8192 were generated. 
Denoting the ``correlation dimension'' of the original data by $f$, 
the mean of the distribution of this
feature for the surrogate data by $ \mu_{surr} $ and its the variance
by $ \sigma^2_{surr} $, the result is displayed as :
\beq \label{zet}
z= \frac{|f - \mu_{surr}|}{\sigma_{surr}} \quad .
\eeq

It was confirmed that the distribution of the feature is 
sufficiently well described by a gaussian distribution. Thus, 
$z$ can be related to a confidence interval, since for 50 realizations 
the $t$-distribution of $(f - \mu_{surr} ) / \sigma_{surr}$
is well approximated by a gaussian distribution and $z=1.96$
corresponds to the 5 \% level of significance.

In general, in power of the test investigations a procedure
different from that outlined above is chosen. For a certain
significance level, e.g.~5\%, and different degrees of violation
of the null hypothesis, numerous realizations, e.g.~1000, of the 
process are generated and the fraction of rejected null hypotheses
is reported. Due to the high computational burden for calculating
the correlation integral, this procedure is not feasible here.
The above procedure has the drawback that the results depend on
the single realization that is used as basis for the surrogates.
We repeated the analysis reported below for independent realizations
and found no qualitative differences for different realizations.

For the first violation of the null hypothesis, we increase 
$ \text{Mod}_{\text{amp}} $ in Eq.(\ref{process1})
from zero, i.e.~no violation, to 0.5 in steps of 0.1.
The distribution of these data are not gaussian 
for $\text{Mod}_{\text{amp}} > 0 $. 
Thus, the amplitude adjusted surrogate data algorithm  
 \cite{theiler92} was applied. The deviation from gaussianity
is weak for the range of violations chosen. We also applied
the algorithm without amplitude adjustment and did not found 
significant different results.

\begin{figure}
\epsfxsize=8.4cm
\epsfbox{  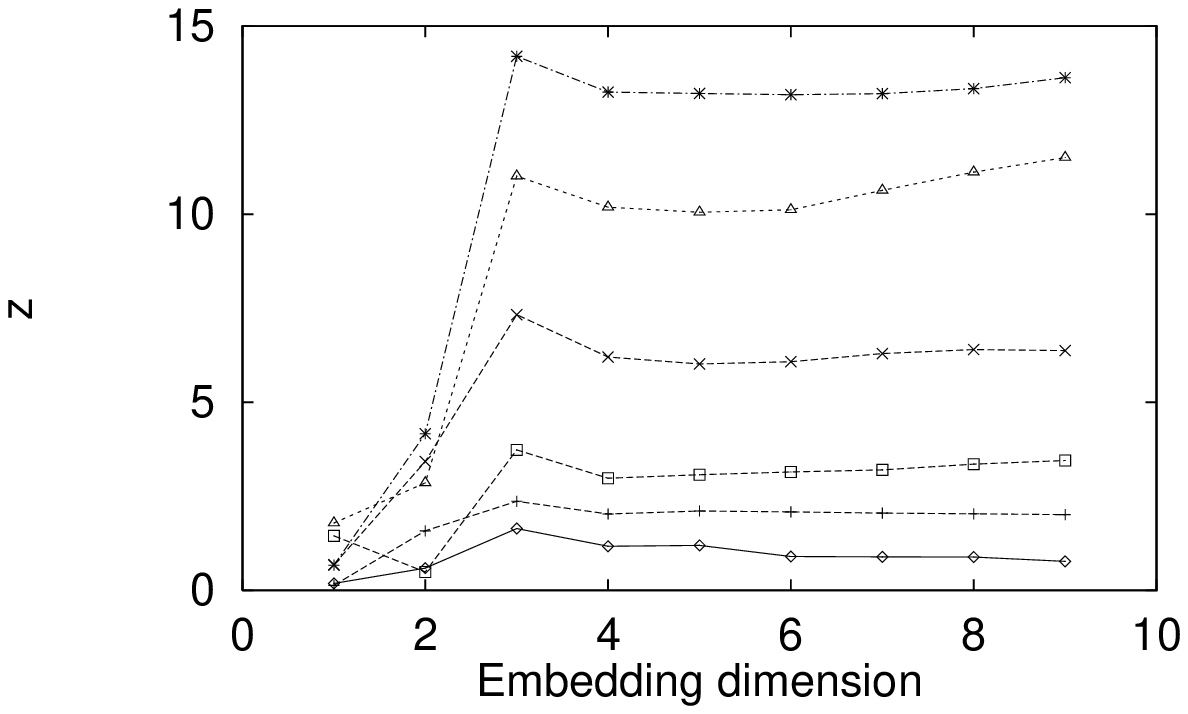   }
\caption{\label{ar2amp_res}Results of the simulation study for
the amplitude modulated process. Shown is $ z $ in
dependence on the embedding dimension for different degrees
$ \text{Mod}_{\text{amp}} $ of violation (  $\Diamond=0$, $+=0.1$,  
$\Box=0.2 $, $\times=0.3$, $\triangle=0.4$ $\ast=0.5$).}
\end{figure}

Fig.~\ref{ar2amp_res} displays the result of the simulation study.
In dependence on the embedding dimension,  $ z $
is displayed for different degrees of violation of the null
hypothesis. 

As expected, without any violation, the $ z $ values
stay within the 2$\sigma$ region given by $ z \le 1.96 $.
A modulation depth $ \text{Mod}_{\text{amp}} $ of $0.1$ and $0.2$ 
leads to results at the border of 5\% significance. Starting from
 $\text{Mod}_{\text{amp}} = 0.3$, see Fig.~\ref{ar2_dat}b, the null
hypothesis is clearly rejected at the 5\% level of significance
whenever the embedding dimension is large enough to reconstruct 
the second order process appropriately.

\begin{figure}
\epsfxsize=8.4cm
\epsfbox{  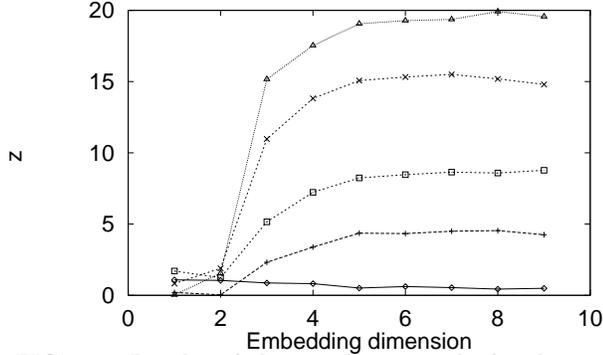 }
\caption{\label{ar2fre_res} Results of the simulation study for
the period modulated process. Shown is $ z $ in
dependence on the embedding dimension for different degrees $ \text{Mod}_T $
of violation ( $\Diamond=0$, $+=1$,  $\Box=1.5 $, $\times=2$, 
$\triangle=3$).}
\end{figure}

To investigates the effect of a variation in the period of the 
linear stochastic process, we increase $\text{Mod}_T$ in 
Eq.(\ref{process2_1},\ref{process2_2}) from zero to three.The 
distribution of these data are gaussian independent from the value 
of $\text{Mod}_T$. Thus, no amplitude adjustment was necessary.
Again, the distribution of the feature is sufficiently well
described by a gaussian distribution.
Fig.~\ref{ar2fre_res} displays the result of the simulation study.

For all degrees of violation, the violation is not detected when
the embedding dimension is too small to unfold the dynamics in phase
space. Otherwise, a modulation of the period of 15 \%, see 
Fig.~\ref{ar2_dat}c, leads to a clear rejection
of the null hypothesis at the 5 \% level of confidence.
\section*{IV. Conclusion}
The simulation studies reported in this paper indicate that surrogate
data testing for linear, stochastic, gaussian stationary processes is
powerful against a violation of the assumption of stationarity. 
Thus, a significant result of the test does not necessarily 
indicate a non-linear or even chaotic process
underlying the data. It might have simply be caused by a
non-stationarity of the process. 



\begin{thebibliography}{10}

\bibitem{may76}
R. May, Nature {\bf 261},  459  (1976).

\bibitem{wolf85}
A. Wolf, J. Swift, H. Swinney, and L. Vastano, Physica D {\bf 16},  285
  (1985).

\bibitem{grassberger83a}
P. Grassberger and I. Procaccia, Physica D {\bf 9},  189  (1983).

\bibitem{sugihara90}
G. Sugihara and R. May, Nature {\bf 344},  734  (1990).

\bibitem{theiler92}
J. Theiler {\it et~al.}, Physica D {\bf 58},  77  (1992).

\bibitem{rapp93b}
P. Rapp, Biologist {\bf 40},  89  (1993).

\bibitem{glass93}
L. Glass and D. Kaplan, Med. Prog. Tech. {\bf 19},  115  (1993).

\bibitem{jedynak94}
A. Jedynak, M. Bach, and J. Timmer, Phys.Rev. E {\bf 50},  1770  (1994).

\bibitem{kantz95}
H. Kantz and T. Schreiber, Chaos {\bf 5},  143  (1995).

\bibitem{schiff94}
S. Schiff {\it et~al.}, Biophys. J. {\bf 67},  684  (1994).

\bibitem{stam95}
C. Stam {\it et~al.}, Electroencephal. clin. Neurophys. {\bf 95},  309  (1995).

\bibitem{mrowka96}
R. Mrowka, A. Patzak, E. Schubert, and P. Persson, Cardiovasc. Res. {\bf 31},
  447  (1996).

\bibitem{guzzetti96}
S. Guzzetti {\it et~al.}, Cardiovasc. Res. {\bf 31},  441  (1996).

\bibitem{yamamoto93}
Y. Yamamoto {\it et~al.}, Biol. Cybern. {\bf 69},  205  (1993).

\bibitem{yip95}
K. Yip, D. Marsh, and N. Holstein-Rathlou, Physica D {\bf 80},  95  (1995).

\bibitem{pradhan96}
N. Pradhan and P. Sadasivan, Phys. Rev. E {\bf 53},  2684  (1996).

\bibitem{osborne89}
A. Osborne and A. Provenzale, Physica D {\bf 35},  357  (1989).

\bibitem{theiler91}
J. Theiler, Phys. Lett. A {\bf 155},  480  (1991).

\bibitem{goldberger90}
A. Goldberger, D. Rigney, and B. West, Sci. Am. {\bf 262},  42  (1990).

\bibitem{garnder90}
W. Gardner, {\em Introduction to random processes with application to signals
  and systems} (MacGraw-Hill, New {Y}ork, 1990).

\bibitem{kaplan97}
D. Kaplan,  in {\em Frontiers of Blood Pressure and Heart Rate Analysis},
  edited by M.~D. et~al. (IOS Press, New York, 1997).

\bibitem{theiler86}
J. Theiler, Phys. Rev. A {\bf 34},  2437  (1986).

\bibitem{theiler93}
J. Theiler and T. Lookman, Int. J. Bif. Chaos {\bf 3},  765  (1993).

\bibitem{pritchard92}
W. Pritchard and D. Duke, Psychophysiology {\bf 29},  182   (1992).

\end{thebibliography}
\end {document}